\documentclass{svjour3}                     % onecolumn (standard format)
\smartqed  % flush right qed marks, e.g. at end of proof
\usepackage{graphicx}
\begin{document}

\title{$\Delta \rho \pi$ interaction leading to $N^*$ and $\Delta^*$ resonances
\footnote{Presented at the 21st European Conference on Few-Body
Problems in Physics, Salamanca, Spain, 30 August - 3 September
2010.} }


\author{Ju-Jun Xie \and A. Mart\'inez Torres  \and E. Oset \and P. Gonz\'alez.
}

\institute{Ju-Jun Xie \at Instituto de F\'\i sica Corpuscular
(IFIC), Centro Mixto CSIC-Universidad de Valencia, Institutos de
Investigaci\'on de Paterna, Aptd. 22085, E-46071 Valencia, Spain \\
Department of Physics, Zhengzhou University, Zhengzhou,
Henan 450001, China \\ \email{xiejujun@ific.uv.es}           %  \\
%             \emph{Present address:} of F. Author
%  if neede
\and A. Mart\'inez Torres \at Yukawa Institute for Theoretical
Physics, Kyoto University, Kyoto 606-8502, Japan \and E. Oset and P.
Gonz\'alez \at Departamento de F\'\i sica Te\'orica and Instituto de
F\'\i sica Corpuscular (IFIC), Centro Mixto CSIC-Universidad de
Valencia, Institutos de Investigaci\'on de Paterna, Aptd. 22085,
E-46071 Valencia, Spain
%\and P. Gonz\'alez \at Departamento de F\'\i sica Te\'orica and
%Instituto de F\'\i sica Corpuscular (IFIC), Centro Mixto
%CSIC-Universidad de Valencia, Institutos de Investigaci\'on de
%Paterna, Aptd. 22085, E-46071 Valencia, Spain
 }

\date{Received: date / Accepted: date}

\maketitle

\begin{abstract}
We have performed a calculation for the three body $\Delta \rho \pi$
system by using the fixed center approximation to Faddeev equations,
taking the interaction between $\Delta$ and $\rho$, $\Delta$ and
$\pi$, and $\rho$ and $\pi$ from the chiral unitary approach. We
find several peaks in the modulus squared of the three-body
scattering amplitude, indicating the existence of resonances, which
can be associated to known $I=1/2, 3/2$ and $J^P=1/2^+, 3/2^+$ and
$5/2^+$ baryon states.

\end{abstract}

\section{Introduction}
\label{intro}

Our knowledge on the baryon resonances mainly comes from $\pi N$
experiments and is still under
debate~\cite{pdg2008,Arndt:2006bf,Arndt:2008zz,Arndt:2009nv}. The
information extracted from photon nucleon reactions have helped in
making progress in this field, reconforming many known resonances
and claiming evidence for new
ones~\cite{Burkert:2004sk,Matsuyama:2006rp,JuliaDiaz:2007kz,Anisovich:2009zy,Horn:2007pp,Klempt:2006sa,Arends:2008zz}.
The fact that some known resonances are explained in terms of three
body systems of two mesons and one
baryon~\cite{MartinezTorres:2007sr,Khemchandani:2008rk} should
certainly stimulate work looking for resonances in three body final
states of reactions. In this sense a suggestion is made
in~\cite{meissner} to look for a predicted state of
$N\bar{K}K$~\cite{Jido:2008kp,MartinezTorres:2008kh} in the $\gamma
p \to K^+ K^- p$ reaction close to threshold.

The main aim of the present work is to investigate the three-body
$\Delta \rho \pi$ system considering the interaction of the three
components among themselves keeping in mind the expected strong
correlations of the $\Delta \rho$ system which generate the $N^*$
and/or $\Delta^*$ bound states. For this purpose, we have solved the
Faddeev equations by using the fixed center approximation(FCA) in
terms of two body $\Delta \pi$ and $\rho \pi$ scattering amplitudes.

The FCA to the Faddeev equations is a tool which has proved to be
efficient and accurate to study the interaction of particles with
bound states of a pair of particles at very low energies, or below
threshold~\cite{Chand:1962ec,Barrett:1999cw,Deloff:1999gc,Kamalov:2000iy}.
Recently, this approach was used in Ref.~\cite{Roca:2010tf} to
describe the $f_2(1270)$, $\rho_3(1690)$, $f_4(2050)$,
$\rho_5(2350)$ and $f_6(2510)$ resonances as multi-$\rho(770)$
states, and in Ref.~\cite{YamagataSekihara:2010qk} to study the
$K^*_2(1430)$, $K_3^*(1780)$, $K^*_4(2045)$, $K^*_5(2380)$, and a
not yet discovered $K^*_6$ resonances as $K^*-$multi$-\rho$ states.
The success of these works encourages us to extend the method to
study the present $\Delta \rho \pi$ system.

\section{Formalism and results}

For the three body $\Delta \rho \pi$ system, we consider $\Delta
\rho$ as a bound state of $N^*$($I_{\Delta \rho}=1/2$) resonance or
$\Delta^*$($I_{\Delta \rho}=3/2$) resonance, which allows us to use
the FCA to solve the Faddeev equations. The external $\pi$ meson
interacts successively with the $\Delta$ baryon and the $\rho$ meson
which form the $\Delta \rho$ cluster. In terms of two partition
functions $T_1$ and $T_2$, the FCA equations are
\begin{eqnarray}
T_1 &=& t_1+t_1G_0T_2 , \label{T1} \\
T_2 &=& t_2+t_2G_0T_1, \label{T2} \\
T &=& T_1+T_2,        \label{T}
\end{eqnarray}
where $T$ is the total three-body scattering amplitude and
$T_i$($i=1,2$) accounts for the diagrams starting with the
interaction of the external particle with the particle $i$ of the
compound system and $t_i$ represents the two body $\Delta \pi$ and
$\rho \pi$ unitarized scattering amplitudes.

Next, we will show the results obtained from the scattering
amplitude of the $\Delta \rho \pi$ system. We evaluate the
scattering amplitude $T$ matrix of Eq.~(\ref{T}) and associate the
peaks of $|T|^2$ to resonances. In table~\ref{tab:results} we show a
summary of the findings obtained from our model and the tentative
association to known states~\cite{pdg2008}.

\begin{table}[htbp]
\caption{The properties of the generated resonances with our model
and their possible PDG counterparts. \label{tab:results}}
\begin{center}
\begin{tabular}{c|c|cccc}
\hline  $I_{\Delta \rho}, I_{total}$ & Mass of our  & \multicolumn{4}{c}{PDG data} \\
\vspace*{-0.3cm}
       &  &  & & &   \\
      & model(MeV)  & name &$J^P$  & mass(MeV)  & status \\
\hline $\frac{1}{2},\frac{1}{2}$ & $\sim 1850$
& $N^*(1900)$ & $3/2^+$ & $  1900$ & ** \\
\hline $\frac{1}{2},\frac{3}{2}$ & $\sim 1800$ &
 $\Delta^*(1750)$ & $1/2^+$ & $ 1750$ & *   \\
 &  &$\Delta^*(2000)^?$ & $5/2^+$ & $1724 \pm 61$ & Ref.~\cite{vrana00}  \\
 &  &$\Delta^*(2000)^?$ & $5/2^+$ & $1752 \pm 32$ & Ref.~\cite{manley92}  \\
 &$\sim 1900$  &$\Delta^*(1905)$ & $5/2^+$ & $1865 - 1915$ & ****  \\
 &  &$\Delta^*(1920)$ & $3/2^+$ & $1900 - 1970$ & ***  \\
 &$\sim 2200$  &$\Delta^*(2000)^?$ & $5/2^+$ & $2200 \pm 125$ & Ref.~\cite{cutkosky80}  \\
\hline $\frac{3}{2},\frac{1}{2}$ & $\sim 2000$
& $N^*(2000)$ & $5/2^+$ & $  2000$ &** \\
\hline $\frac{3}{2},\frac{3}{2}$ & $\sim 2000$
& ? &?  &? &? \\
\hline

\end{tabular}
\end{center}
\end{table}

\section{Discussions and Conclusions}

We have performed a Faddeev calculation for the three body $\Delta
\rho \pi$ system by using the fixed center approximation, taking the
interaction between $\Delta$ and $\rho$, $\Delta$ and $\pi$, and
$\rho$ and $\pi$ from the chiral unitary approach. The $\Delta \rho$
interaction within the framework of the hidden-gauge formalism in
$I=1/2$ sector describes the $N^*(1675)J^P=5/2^-$ as a $\Delta \rho$
bound state, then we write the three-body interaction in terms of
two-body ($\Delta \pi$ and $\rho \pi$) $s-$wave scattering
amplitudes based on the chiral Lagrangians. The three body states
found are degenerated in $J^P=1/2^+,3/2^+,5/2^+$. We found
candidates in the PDG book which can be associated to the states
obtained, but one of them, with isospin $3/2$ and mass around $2000$
MeV, is missing. It is very interesting to observe that, even if the
$\Delta \rho \pi$ system allows for $I=5/2$, the dynamics of the
system precludes the formation of these exotic states.

\bigskip
\noindent

\begin{acknowledgements}
This work is partly supported by DGICYT Contract No. FIS2006-03438,
the Generalitat Valenciana in the project PROMETEO and the EU
Integrated Infrastructure Initiative Hadron Physics Project under
contract RII3-CT-2004-506078. Ju-Jun Xie acknowlwdges Ministerio de
Educaci\'on Grant SAB2009-0116. The work of A.~M.~T.~is supported by
the Grant-in-Aid for the Global COE Program ``The Next Generation of
Physics, Spun from Universality and Emergence" from the Ministry of
Education, Culture, Sports, Science and Technology (MEXT) of Japan.
\end{acknowledgements}

%\begin{acknowledgements}
%\end{acknowledgements}



\end{document}